\documentclass[pre,nofootinbib,superscriptaddress,onecolumn,preprintnumbers]{revtex4}
\usepackage{graphicx}
\usepackage{amsmath,amssymb}
\usepackage{color}
\newcommand{\leri}[1]{\left(#1\right)}
\newcommand{\lerisq}[1]{\left[#1\right]}

\begin{document}
\title{On the Effects of Tokamak Plasma Edge Symmetries on Turbulence Relaxation}

\author{N. Carlevaro}
\email{nakia.carlevaro@enea.it}
\affiliation{ENEA, Fusion and Nuclear Safety Department, C. R. Frascati,\\ Via E. Fermi 45,  Frascati, 00044 Roma, Italy}

\author{G. Montani}
\email{giovanni.montani@enea.it}
\affiliation{ENEA, Fusion and Nuclear Safety Department, C. R. Frascati,\\ Via E. Fermi 45,  Frascati, 00044 Roma, Italy}
\affiliation{Physics Department, ``Sapienza'' University of Rome, \\ P.le Aldo Moro 5, 00185 Roma, Italy}

\author{F. Moretti}
\email{fabiomoretti2718@gmail.com}
\affiliation{ENEA, Fusion and Nuclear Safety Department, C. R. Frascati,\\ Via E. Fermi 45,  Frascati, 00044 Roma, Italy}

\begin{abstract}
The plasma edge of a tokamak configuration is characterized by turbulent dynamics leading to enhanced transport. We construct a simplified 3D Hasegawa--Wakatani model reducing to a single partial differential equation for the turbulent electric potential dynamics. Simulations demonstrate how the 3D turbulence relaxes on a 2D axisymmetric profile, corresponding to the so-called interchange turbulence. The spectral features of this regime are found to be strongly dependent on the initialization pattern. We outline that the emergence of axisymmetric turbulence is also achieved  when the corresponding mode amplitude is not initialized. Then, we introduce the symmetries of the magnetic X-point of a tokamak configuration. We linearize the governing equation by treating the poloidal field as a small correction. We show that it is not always possible to solve the electric potential dynamics following a perturbative approach. This finding, which is due to resonance between the modes of the background and the poloidal perturbation, confirms that the X-point symmetries can alter the properties of turbulent transport in the edge region.
\end{abstract}

\maketitle

\section{Introduction}
One of the most important ingredients characterizing the nature of the heat and particle transport in a tokamak device consists of the turbulent dynamics of the electric field in the edge region and, in particular, close to the X-point \cite{scott07,tcv21,stegmeir-GRILLIX-19}. Since, in the tokamak edge region, the plasma has a significant collisional character, and the turbulence scale is super Debye-sized, we can represent the plasma dynamics via a two-fluid model (ions and electrons) for which the current density vector is divergence-free. 

In this framework, the plasma turbulent transport is mainly associated with the so-called nonlinear drift response \cite{scott90,scott02,scott07}, and the corresponding basic coupling between the dynamics of the different variables is due to the parallel (to the background magnetic field) current. The basic features of the electrostatic turbulence (i.e., the magnetic fluctuations are neglected for sufficiently large values of the plasma's $\beta$ parameter) are captured by the so-called Hasegawa--Wakatani model \cite{hase-waka83,hase-waka87,hase-mima18}, which corresponds to a coupling between the dynamics of the number density and the electric vorticity (i.e. the Laplacian of the electric potential). In two dimensions, when the tokamak axisymmetric structure is extended to the turbulent fluctuations, the electric potential field dynamics is isomorphic to that of the stream function in the Euler equation for an incompressible fluid \cite{seyler75,montani-fluids2022,montani-physicad-2022}. However, in such a limit, if the magnetic field is assumed to be along the $z$ direction, then the drift coupling is absent, and the steady state of the system is associated with an enstrophy cascade and, in some cases, also responsible for an inverse energy cascade (known as the ``condensation phenomenon''), which allows for the formation of large-scale eddies \cite{kraichnan80}. Studies on the properties of the three-dimensional turbulence relaxation on a two-dimensional profile were discussed in \cite{biskamp95,montani-physicad-2022} but were for when the axisymmetric mode was always initialized in the numerical analysis. 

In this respect, in this work, we consider a reduced model equivalent to a Hasegawa--Wakatani scenario but assume that the linear drift instability trigger is negligible (i.e., treating the background plasma density as a constant parameter). In this context, we simulate the dynamics for when the magnetic field of the background configuration is taken along the $z$ axis and the axisymmetric fluctuation is not initialized. Our study demonstrates that the three-dimensional turbulence is destined to decay, and apart from the viscous damping, all the energy is transferred, exciting the axisymmetric mode, which manifests a spectral feature compatible with a constant energy distribution which would correspond to a two-dimensional initialization with a random amplitude.

Furthermore, we study the properties of the linearized plasma dynamics when the X-point geometry is considered \cite{snowflake15} so that a parallel derivative is now also present in the axisymmetric case, which is the object of our investigation in light of the previous analyses on the 3D case. As a novel result of this study, we see that if we treat the poloidal magnetic field as a small perturbation, then a separation of the linearized equation is possible, and the stability properties can be properly discussed. More precisely, we see that when the perturbative approach remains valid, the stability of the dynamics is ensured. However, in the more external region of the edge plasma, the perturbation scheme fails. This is a consequence of a secular relative increase in the perturbation with respect to the background evolution. The two terms become of the same order in a time scale shorter than the damping rate. This result clearly suggests that, although close enough to the X-point, the poloidal magnetic field is very small in comparison with the toroidal contribution, but its presence can deeply influence the linear and nonlinear stability properties of the plasma because it cannot always be treated on a perturbative level.

This paper is organized as follows. In Section \ref{sec2}, we start by describing the physical scheme and the fundamental hypotheses we address in the paper. The 3D reduced model for turbulence is derived. In Section \ref{sec3vs2}, a closed equation describing the dynamics of the electric potential fluctuations is discussed when the magnetic field is assumed to be along the toroidal direction. We develop the numerical simulations of the reduced model by implementing a truncated Fourier approach. We neglect nonideal effects. The system dynamics is evolved until the thermal equilibrium is reached, and we show how the axisymmetric mode behaves as an attractor. In Section \ref{seclin}, we assume a nonzero poloidal magnetic field, and we analyze, by resorting only to analytical techniques, the linearized equation ruling the reduced 2D dynamics also having the form of a single equation in this case. We show that a perturbative approach in the construction of regular solutions is not always feasible. This is due to a resonance phenomenon which can lead to secular growth for low-wavenumber modes, according to the specific values of the physical parameters characterizing the system. Our concluding remarks follow.

\section{3D Turbulence Reduced Model}\label{sec2}
\begin{figure}[ht!]
\includegraphics[width=10cm]{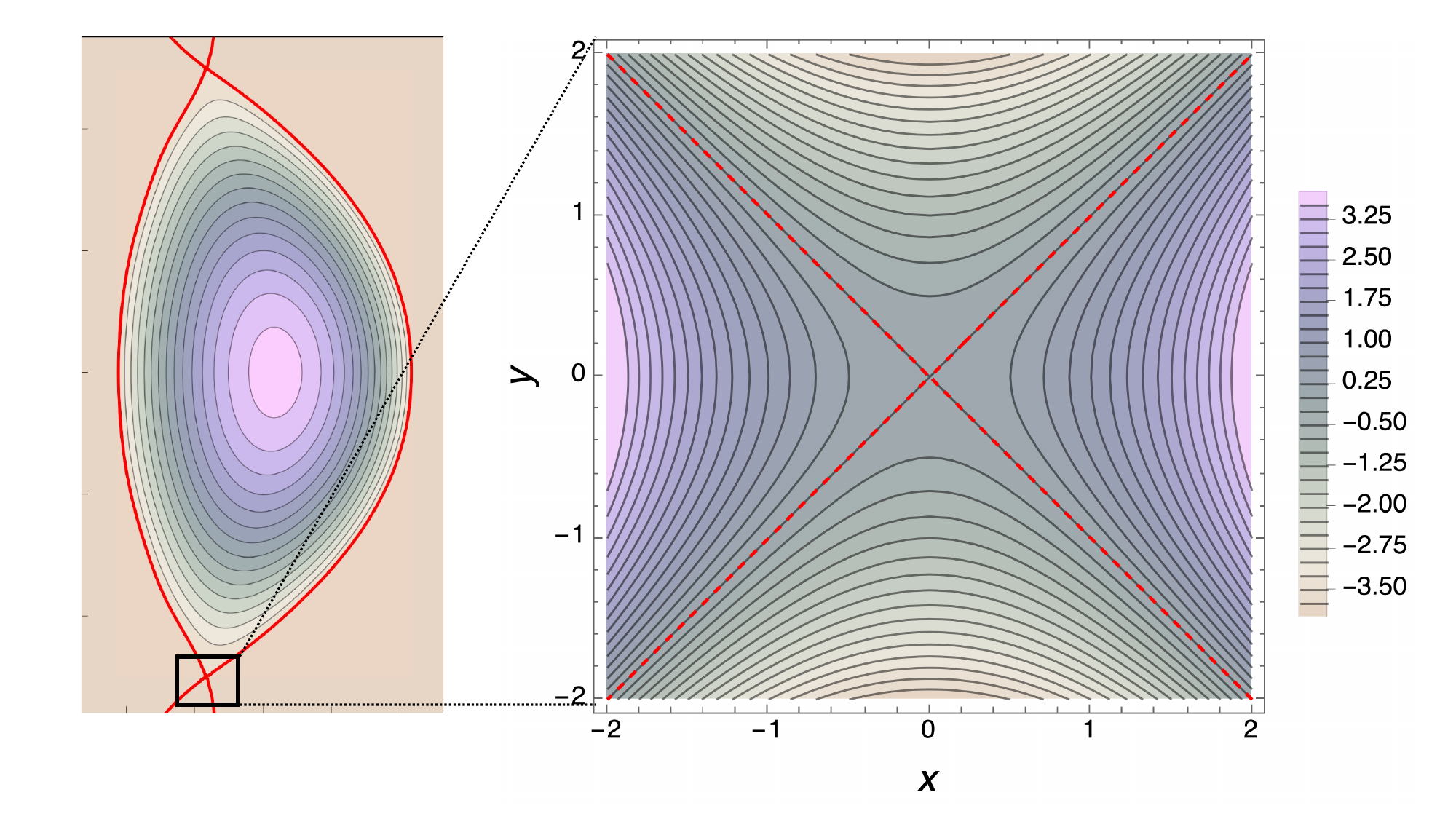}
\caption{Representation of a tokamak double-null poloidal magnetic configuration and a zoomed-in section with the contour plot, in arbitrary units, of $\psi\propto (x^2-y^2)$. The red line represents the separatrix (the last closed magnetic surface) taken at $\psi=0$ (color online).}
\label{figXP}
\end{figure}

In this section, we construct the reduced model for the turbulent dynamics representing a standard scheme for the nonlinear low-energy drift response, which can be recast in a single equation for the electric fluctuations. We start by addressing the morphology properties of the local equilibrium corresponding to the background of our dynamical perturbative approach. Specifically, we consider a small poloidal region close to the X-point of a tokamak configuration described by the Cartesian coordinates $(x,y,z)$ centered in the null, thus neglecting the effects of toroidal curvature. The magnetic field $\textbf{B}$ can be expressed via the magnetic flux function $\psi(x,y)$ as
\begin{equation}
\textbf{B} = -\partial_y\psi\hat{\textbf{e}}_x + 
	\partial_x\psi\hat{\textbf{e}}_y + B_t\hat{\textbf{e}}_z
	\, , 
	\label{lm}
\end{equation}
where $B_t=const.$ can be thought as the toroidal magnetic component of a tokamak and is taken as the dominant contribution. Under the assumption of dealing with a low-density and sufficiently cold plasma, we impose the Amp\`ere law \cite{snowflake15} to the flux function (i.e., $\partial^2_x\psi + \partial^2_y\psi = 0 $), obtaining $\psi = (x^2 - y^2)B_{p}/2$. Here, $B_p$ is of the dimensions $B$/length. In Figure \ref{figXP}, we present a qualitative sketch of the poloidal plane of a typical double-null tokamak configuration together with a zoomed-in section representing the zone close to an X-point implementing the above mentioned form of the magnetic flux function in arbitrary units.

By introducing the directional versor $\!\!\hat{\,\,\textbf{b}}$, the background magnetic field in Equation (\ref{lm}) can be rewritten as 
\begin{equation}
\textbf{B}=B\;  \!\!\hat{\,\,\textbf{b}}\,,\qquad
B(x,y)=B_t\sqrt{1+B_p^2(x^2+y^2)/B_t^2}\,,\qquad
\!\!\hat{\,\,\textbf{b}}=\frac{B_p y}{B} \hat{\textbf{e}}_x + \frac{B_p x}{B}\hat{\textbf{e}}_y + \frac{B_t}{B}\hat{\textbf{e}}_z
	\, .
	\label{lm3a}
\end{equation}
We remark that we identify the parallel and perpendicular directions of the dynamics using the magnetic field versor $\!\!\hat{\,\,\textbf{b}}$. This provides specific expressions for the operator $\boldsymbol{\nabla}\equiv \boldsymbol{\nabla}_{\perp}+\boldsymbol{\nabla}_{\parallel}$, where $\boldsymbol{\nabla}_{\parallel}=\!\!\hat{\,\,\textbf{b}}(\!\!\hat{\,\,\textbf{b}}\cdot\boldsymbol{\nabla})$ and $\boldsymbol{\nabla}_{\perp}=\boldsymbol{\nabla}-\boldsymbol{\nabla}_\parallel$, which will be introduced in detail below.

In this work, we consider a homogeneous and isotropic hydrogen-like plasma, and we adopt a two-fluid model. The following assumptions are implemented \cite{hase-mima18,montani-fluids2022}: plasma quasi-neutrality (since the turbulence scale is much larger than the Debye length ) (i.e., $\mathcal{N}_i=\mathcal{N}_e\equiv \mathcal{N}$, with $\mathcal{N}_{i/e}$ indicating the ion/electron number density), and thus we also consider $p_i=p_e=p=\mathcal{N} K_B T$ (where the temperature $T$ is equal for ions and electrons, in which $K_B$ is the Boltzman constant); the ion polarization drift velocity is taken as the only relevant contribution affecting the ortogonal current density; we neglect the diamagnetic effects (i.e., the pressure gradient effects on the electron and ion velocities); and the parallel ion velocity is negligible. 

Finally, as the relevant hypothesis of the model, in the following we assume neglecting of the spatial gradients of the background density by addressing a background characterized by $\mathcal{N}=const.$, $p=const.$ and $T=const.$ Thus, in this work, all the dynamical quantities are perturbations denoted by a $\delta$ symbol.

Let us start by writing down the continuity equation under the hypotheses above, which is the same for electrons and ions by virtue of the charge conservation: 
\begin{equation}
	\frac{d \delta \mathcal{N}}{dt} 
    -\mathcal{D}\nabla^2_{\perp} \delta \mathcal{N} = 
	\frac{1}{e}\boldsymbol{\nabla}_{\parallel}\cdot \delta\textbf{J}_{\parallel}
	\,.
	\label{lm23}
\end{equation}
Here, we have neglected the parallel ion velocity and introduced a diffusion coefficient $\mathcal{D}$ able to phenomenologically model distinct transport regimes \cite{tokam3x16}. Moreover, $\textbf{J}_\parallel$ denotes the parallel current density, with $e$ indicating the electron charge, and we adopt Gaussian units. The Lagrangian derivative $d/dt = ( \partial_t+ \delta\textbf{v}_{\perp}^{e}\cdot\boldsymbol{\nabla})$ is defined by means of the electron $\textbf{E}\times \textbf{B}$ drift velocity:
\begin{align}
	&\delta\textbf{v}_{\perp}^{e} = \frac{c}{B^2} \textbf{E}\times\textbf{B} = \frac{c}{B}\;\!\!\hat{\,\,\textbf{b}}\times \boldsymbol{\nabla}_\perp\delta\phi\,=\nonumber\\
 &=\frac{c}{B^2}\left( 
 (B_p x\partial_z\delta\phi-B_t\partial_y\delta\phi)\hat{\textbf{e}}_x+
 (B_t\partial_x\delta\phi-B_p y\partial_z\delta\phi)\hat{\textbf{e}}_y+
 (B_p y\partial_y\delta\phi-B_p x\partial_x\delta\phi)\hat{\textbf{e}}_z\right)\,,
	\label{lm12}
\end{align}
where $c$ denotes the light speed, while $\delta\phi$ is the fluctuating electric field potential. When neglecting the parallel components of the viscous stress, the fluctuating ion perpendicular velocity $\delta\textbf{v}_{\perp}^{i}$ obeys the following equation:
\begin{equation}
\frac{d\,\delta\textbf{v}_{\perp}^{i}}{dt} = 
	 \frac{e}{m_i}\left( -\boldsymbol{\nabla}_{\perp}\delta\phi + 
	\delta\textbf{v}_{\perp}^{i}\times \textbf{B}/c\right) 
	+ \nu \nabla^2_{\perp}\,\delta\textbf{v}_{\perp}^{i}
	\, , 
 \label{lm13}
\end{equation}
where $\nu$ is the specific ion viscosity and $m_i$ is the ion mass. By implementing the assumption $\mathcal{N}_i=\mathcal{N}$, the (constant) viscosity coefficient assumes the following form \cite{wesson,hase-waka83}:
\begin{align}
    \nu=\frac{1}{m_i\mathcal{N}}\frac{(3/10)\;\mathcal{N} K_B T}{\Omega_i^2 \tau_{ii}}\,,
    \qquad\qquad
    \tau_{ii}=\frac{3\sqrt{\mathcal{N}}(K_B T)^{3/2}}{4\sqrt{\pi}\;e^4 \mathcal{N} \mathrm{ln}\Lambda_{ii}}\,,
    \label{visc}
\end{align}
where we set $\mathrm{ln}\Lambda_{ii}=21$ and we introduced the ion gyrofrequency $\Omega_i=eB_t/c m_i$ related to $B_t$ (the ion Larmor radius results $\rho_i^2=K_B T/m_i \Omega_i^2$). Such a frequency represents, in general, an upper bound for turbulence, and we can thus safely assume  $\delta\textbf{v}_{\perp}^{i} = \delta\textbf{v}_{\perp}^{e} + \tilde{\textbf{v}}_{\perp}^{i}$, with $\tilde{\textbf{v}}_{\perp}^{i}$ being a small correction. The leading order of Equation (\ref{lm13}) thus provides the expression for the correction $\tilde{\textbf{v}}_{\perp}^{i}$:
\begin{equation}
	\tilde{\textbf{v}}_{\perp}^{i}=\frac{cm_i}{Be}\Big( 
	\frac{d\,}{dt} - \nu\nabla^2_{\perp}\Big)(\,\!\!\hat{\,\,\textbf{b}}\times \delta\textbf{v}_\perp^e) =-\frac{c^2m_i}{Be}\Big( 
	\frac{d\,}{dt} - \nu\nabla^2_{\perp}\Big)\frac{1}{B} \boldsymbol{\nabla}_\perp\delta\phi
	\,.
	\label{lm16}
\end{equation}
In this scheme, the orthogonal current is $\delta\textbf{J}_{\perp} \equiv \mathcal{N} e(\delta\textbf{v}_{\perp}^{i} -\delta\textbf{v}_{\perp}^{e})=\mathcal{N} e \,\tilde{\textbf{v}}_{\perp}^{i}$. Thus, the charge conservation $\boldsymbol{\nabla}\cdot\delta\textbf{J}=0$ can be recast as
\begin{equation}
\boldsymbol{\nabla}_\perp\cdot\left(
\frac{1}{B}\frac{d\,}{dt}\frac{1}{B} \boldsymbol{\nabla}_\perp \delta\phi\right)- \nu
\boldsymbol{\nabla}_\perp\cdot\left(\frac{1}{B}\nabla_\perp^2\frac{1}{B} \boldsymbol{\nabla}_\perp \delta\phi\right)=
\frac{1}{\mathcal{N}c^2m_i}\boldsymbol{\nabla}_{\parallel}\cdot\delta\textbf{J}_{\parallel}\,, 
	\label{lm18} 
\end{equation}

Let us now discuss the parallel electron momentum balance. If we take into account only a constant parallel conductivity coefficient $\sigma\equiv1.96\, \mathcal{N} e^2/m_e \nu_{ei}$, where $\nu_{ei}$ is the electron-ion collision frequency (with $m_e$ being the electron mass), the momentum balance reads as follows: 
\begin{align}
\delta\textbf{J}_{\parallel} = \frac{\sigma}{\mathcal{N}e} \boldsymbol{\nabla}_{\parallel}\delta p -\sigma \boldsymbol{\nabla}_{\parallel}\delta\phi\,, \label{eqj}
\end{align}
where, according to the assumptions described above, we can link the pressure and density using $\delta p = K_B T\delta \mathcal{N}$.

Let us move to the dimensionless coordinates $\tau\equiv \Omega_i t$, $u\equiv(2\pi/L_p) x$, $v\equiv(2\pi/L_p) y$, $w\equiv (2\pi/L_t) z$ (where $L_p$ and $L_t$ are two spatial scales with $L_t\gg L_p$). We thus find $(w,u,v)$ running in $[0,2\pi)$. We also introduce the parameter
\begin{align}
\epsilon=\frac{B_p}{B_t}\,\Big(\frac{L_p}{2\pi}\Big)\,,
\end{align}
thus gaining from Equation (\ref{lm3a})
\begin{align}
B= \gamma B_t\,,\qquad
\gamma(u,v)=\sqrt{1+\epsilon^2(u^2+v^2)}    
\end{align}
In this scheme, the dimensionless parallel gradient $\textbf{D}_{\parallel}=(L_p/2\pi)\boldsymbol{\nabla}_{\parallel}$ reads as follows:
\begin{align}\label{ml23}
\textbf{D}_{\parallel}=\Big( 
\frac{\epsilon}{\gamma}(v\partial_u+u\partial_v)+ \frac{L_p/L_t}{\gamma}\,\partial_w
\Big)\!\!\hat{\,\,\textbf{b}}\,,
\end{align}
from which the orthogonal gradient $\textbf{D}_{\perp}$ and the Laplacian operators $D_{\parallel}^2$ and $D^2_{\perp}$ can be easily derived. Here, for the sake of simplicity, we do not write down their explicit forms, but we directly introduce the reduced expressions in the following sections.

By setting $\Phi\equiv e\,\delta\phi /K_BT$, $\bar{\mathcal{N}}=\delta\mathcal{N}/\mathcal{N}$ and $\textbf{Y}_{\parallel} =(2\pi/L_p)\delta\textbf{J}_{\parallel}/ \mathcal{N}e\Omega_i$, \mbox{Equations (\ref{lm23}) and (\ref{lm18})} can be rewritten as 
\begin{align}
\frac{d}{d\tau}\bar{\mathcal{N}}-\bar{\mathcal{D}} D^2_{\perp} \bar{\mathcal{N}} &= 
\boldsymbol{D}_{\parallel}\cdot\textbf{Y}_{\parallel}\;,\label{eq1}\\
\alpha_1\boldsymbol{D}_\perp\cdot\left(
\frac{1}{\gamma}\frac{d\,}{d\tau}\frac{1}{\gamma}\boldsymbol{D}_\perp \Phi\right)
-\alpha_1\alpha_2
\boldsymbol{D}_\perp\cdot\left(\frac{1}{\gamma}D_\perp^2\frac{1}{\gamma}\boldsymbol{D}_\perp \Phi\right)&= 
\boldsymbol{D}_{\parallel}\cdot\textbf{Y}_{\parallel}\;,\label{eq2}
\end{align}
while, from Equation (\ref{eqj}), we obtain
\begin{equation}
\boldsymbol{D}_{\parallel}\cdot\textbf{Y}_{\parallel} =\alpha_3 (D_{\parallel}^2\bar{\mathcal{N}}  -D_{\parallel}^2\Phi)\,.
\label{lm19}
\end{equation}
Here, we have defined the following dimensionless constants:
\begin{align}
\alpha_1= \rho_i^2\Big(\frac{2\pi}{L_p}\Big)^2\,,\quad
\alpha_2=\frac{\nu}{\Omega_i}\Big(\frac{2\pi}{L_p}\Big)^2\,, \quad
\alpha_3=\frac{v_A^2 \rho_i^2}{\eta_B \Omega_i}\Big(\frac{2\pi}{L_p}\Big)^2\,,\quad
\bar{\mathcal{D}}=\frac{\mathcal{D}}{\Omega_i}\Big(\frac{2\pi}{L_p}\Big)^2\,,
\end{align}
where we have used the notation $v_A=B_t/\sqrt{4\pi \mathcal{N} m_i}$, indicating the Alfv\'en velocity constructed with the toroidal magnetic field, and $\eta_B=c^2/4\pi\sigma$ denotes the magnetic diffusivity. 

The constructed simplified model is depicted in Equations (\ref{eq1}) and (\ref{eq2}) and corresponds to a reduction in the Hasegawa--Wakatani scheme for the plasma turbulence \cite{hase-waka83}. The relevant assumption is related to neglecting the background number of density spatial gradients.

\subsection*{Relevance for Tokamak Physics}

We now discuss to which extent the present model is, in practice, relevant for describing the edge plasma physics of a tokamak machine. First of all, we observe that the two-fluid 
description adopted above is appropriate to 
describe the edge plasma dynamics 
because, immediately out of the separatrix, the 
lower temperature makes the mean free path 
of ions and electrons much smaller than the 
magnetic connection length \cite{tokam3x16,chirkov,sdvizhenskii2021}. Therefore, 
the collisionality increases enough with respect to the core of a tokamak plasma that a fluid model 
is viable, and the kinetic and gyrokinetic effects remain 
small for as far as we are considering turbulence 
phenomena (for which the time scale is much longer than 
the inverse of the ion gyrofrequency). 
In other words, a significant range of the 
operation parameters for present day and incoming tokamak devices \cite{DTT_2021} is associated to physical conditions 
addressed well by a low-frequency drift fluid model, such as the one traced above \cite{scott07}. 
Furthermore, we stress that the typical spatial 
scales of the turbulence phenomena in a tokamak are from few millimeters up to 10 centimeters. Thus, such processes are typically on 
a super Debye spatial scale so that the quasi-neutrality assumption is well posed. 

Finally, we stress that we consider here only 
electrostatic turbulence in the sense that we 
neglect the possible magnetic fluctuations due to 
a nonzero parallel vector potential contribution 
(see \cite{scott02}). It is well known that the 
electrostatic turbulence is able to capture the 
dominant features of heat and particle transport in the tokamak edge plasma when the $\beta$ plasma parameter is 
sufficiently low. In fact, for many operation 
regimes of present day tokamaks, the magnetic 
pressure is a dominant contribution in the plasma 
equilibrium with respect to the thermal phenomena, 
and this fact guarantees a sufficient freezing out 
of the magnetic turbulent fluctuations.

The considerations above state that the so-called Hasegawa--Wakatani model \cite{hase-waka83} is a valuable 
tool for capturing the basic features of the 
tokamak edge turbulent transport, as is also 
consolidated in the literature. However, 
in principle, other smaller effects can be included to make  the picture of the edge physics more detailed, 
like diamagnetic or magnetic curvature contributions.
Evaluating in which physical situations deviations from the Hasegawa--Wakatani scenario need to be accounted for is out of the scope of the present manuscript, and 
it has been widely discussed in the literature \cite{scott07,hase-mima18}.
Instead, we now provide a brief discussion concerning 
the predictivity of our reduced model for the 
tokamak edge dynamics. 

With respect to a standard Hasegawa--Wakatani model, 
in order to deal with a single equation for the electric field, 
we simply neglect the background density gradient, which is 
responsible for the linear triggering of 
the drift instability. Actually, this assumption is justified 
by previous analyses \cite{scott02}, which clarified how the 
nonlinear drift response (being the basic tool to 
describe the edge plasma turbulence) is an intrinsic 
nonlinear self-sustained phenomenon. In other words, 
the emergence of a fully developed drift turbulence regime is 
even favored by the smallness of the linear triggering, 
since the nonlinear interaction of small drift-like fluctuations is sufficient to guarantee the onset of turbulence.
Thus, we can reliably claim that the present reduced model 
has significant physical content, since it offers 
a genuine representation of the basic features of 
the turbulent transport in the tokamak's edge. 
In this sense, the proposed paradigm is the most simple 
but exhaustive representation of a drift fluid nonlinear response. 

A separate discussion deserves the introduction of 
a local geometry for the background magnetic X-point
concerning the linearized system analysis. 
The proposed representation of the background magnetic configuration describes the plasma region immediately 
outside the separatrix and laying in the proximity 
of the null of the poloidal field component well. 
The size of the region we investigate near the 
X-point has been fixed according typical values 
of the operation of the incoming DTT tokamak \cite{DTT_2021}, 
and it corresponds to a small portion of the plasma region 
between the null configuration and the machine walls. 
Actually, here, the plasma--wall interaction has not 
been considered in detail, but it can be figured out as an effective 
increase in the dissipation phenomena.
We also remark that the available spatial
scales we can deal with in the considered model must 
not exceed this mentioned depth, but they must remain 
much greater than the ion's Larmor radius; otherwise, 
the low-frequency drift fluid approximation addressed here 
is no longer fully satisfied.

The analysis is restricted to the X-point region 
because a significant magnetic shear is 
present, and this makes the physical 
and dynamical properties of the electrostatic turbulence quite peculiar. 
The predictivity 
of global turbulent codes (like TOKAM3X and GRILLIX) 
in the tokamak edge region  is currently under  investigation, and the most important 
deviations in comparison to the data just 
concern the region near the divertor \cite{tcv21} and close 
to the null configuration \cite{galassi_torpex_2022}.

Here, we concentrate our attention on the linearized 
dynamics in order to give some physical insights on 
how the X-point morphology can affect the linear 
stability of the background configuration. 
This is a valid starting point for which generalization to the nonlinear dynamics will shed light on the 
peculiar nature of the turbulence close to a null point and how its presence affects the global profile 
of the heat and particle transport, a perspective 
of basic interest in the development of 
efficient fusion devices. In fact, turbulent 
transport is considered the most relevant ingredient 
in generating the so-called anomalous transport \cite{Braginskii65,stangeby}, still remaining the most 
concrete obstacle to reach stable plasma 
configurations for self-sustained fusion reactions.

\section{Reduced Turbulence Simulations in the Presence of a Pure Toroidal Field}\label{sec3vs2}

In this section, we address a pure constant toroidal magnetic field taken along $w$ (i.e., $B_p=0$ and $B=B_t=const$). This implies setting $\epsilon=0$ ($\gamma=1$) in the scheme previously described, and simplified expressions for the Laplace operators are derived from Equation (\ref{ml23}):
\begin{align}\label{lapl1}
D^2_\parallel \to (L_p/L_t)^2\;\partial_w^2\,,\quad
D^2_\perp \to \partial_u^2+\partial_v^2\,.
\end{align}
Also the (normalized) Lagrangian derivative expressed by means of Equation (\ref{lm12}) takes the following simplified form:
\begin{align}
\frac{d}{d\tau}=\partial_{\tau}+\alpha_1(\partial_{u}\Phi \partial_{v}-\partial_{v}\Phi\partial_{u})\,.
\end{align}

Equations (\ref{eq1}) and (\ref{eq2}) can be reduced to a single evolutive equation for the vorticity $D_\perp^2\Phi$. In fact, when comparing the two equations, the constitutive relation can easily be obtained:
\begin{align}
\bar{\mathcal{N}}=\alpha_1 D_\perp^2\Phi\;,
\label{nphi}   
\end{align}
provided that the diffusion coefficient is set to $\mathcal{D}=\nu$. In this scheme, Equation (\ref{eq2}) can be rewritten as
\begin{align}
\partial_{\tau}D^2_{\perp}\Phi +\alpha_1\big( \partial_{u} \Phi \partial_{v} D_{\perp}^2\Phi - \partial_{v}\Phi\partial_{u}D_{\perp}^2\Phi\big)=\alpha_2\; D_{\perp}^4\Phi + \alpha_3\; D_{\parallel}^2D_{\perp}^2\Phi-\alpha_4\;D^2_{\parallel}\Phi \, , 
\label{eqphia}
\end{align}
where $\alpha_4=\alpha_3/\alpha_1= v_A^2/\Omega_i\eta_B$. We stress that in this equation, the first term on the left hand side corresponds to the time evolution of the vorticity (Laplacian of the electric field), while the second one is a pure advection term, in which nonlinearity is the basic ingredient of the 2D interchange-like turbulence. The term on the right hand side is due to the presence of ion viscosity when the expression of the $\textbf{E}\times\textbf{B}$ drift is considered. The last two terms are noted more than the drift coupling term once it is expressed via the parallel electron force balance (i.e., Equation (\ref{lm19})) and the constitutive relation between the vorticity and the number density (i.e., Equation (\ref{nphi})). Our model is a 3D formulation as the Hasegawa--Wakatani model, and we aim just at investigating what
the steady fate of the turbulent transport is when
the non-axisymmetric modes (no longer excited by the linear trigger neglected here) are properly initialized. We will see below via a numerical analysis how the axisymmetric mode becomes an attractor for the turbulent dynamics so that
the pure drift coupling term on the right hand side of Equation (\ref{eqphia}) is
ruled out due to its decaying over time.
The relic's turbulent dynamics thus has a 2D
nature, and the spectrum naturally induces the
corresponding number's density behavior as
discussed in \cite{montani-physicad-2022}.

\begin{figure}[ht!]
\includegraphics[width=4.5cm]{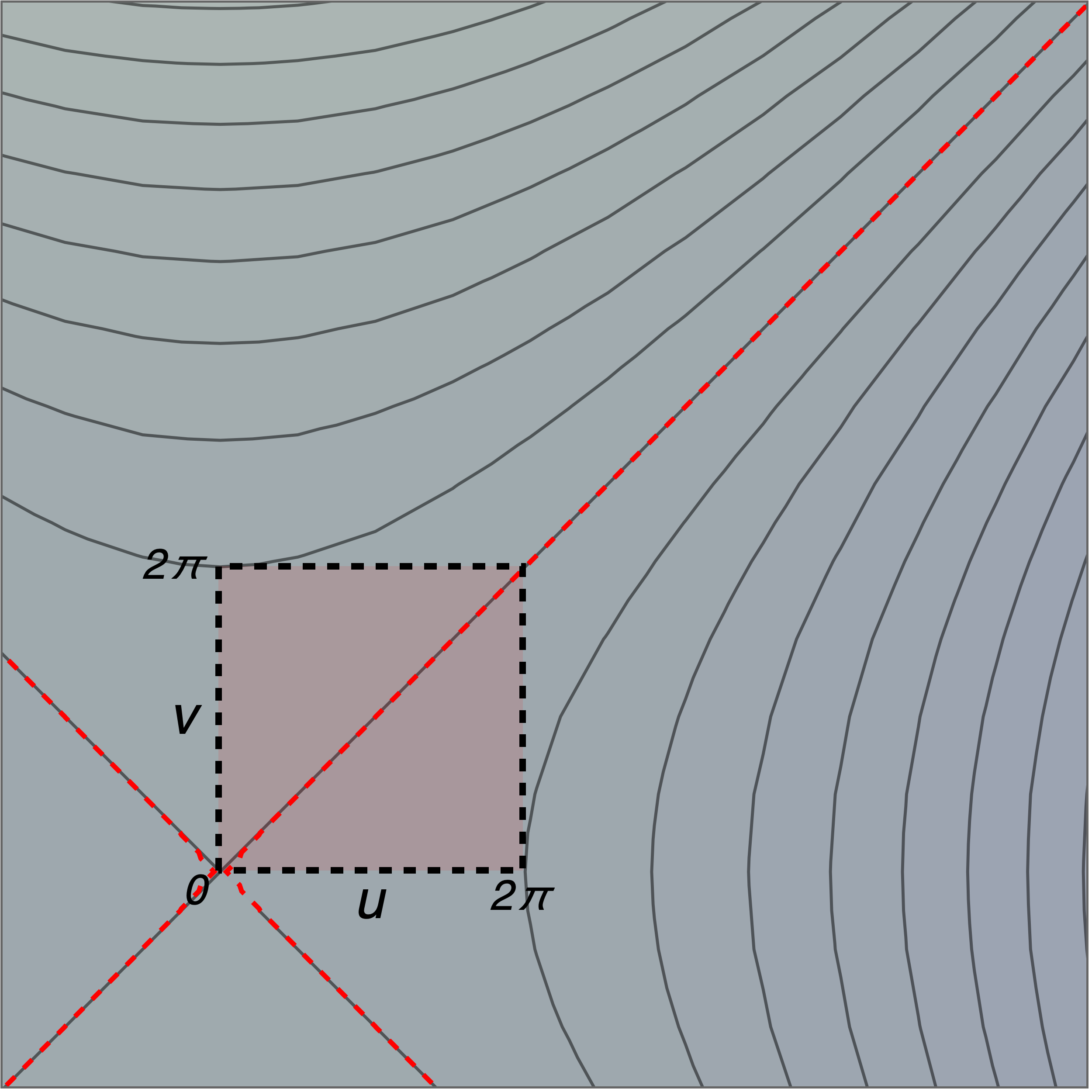}
\caption{Qualitative scheme of the poloidal box for simulations where we are assuming $B_p=0$ (color online).}
\label{figLp}
\end{figure}

We now study a small plasma region close to the X-point. In Figure \ref{figLp}, we show a sketch of the poloidal box we are considering (compare with Figure \ref{figXP}) and where the assumption $B_p=0$ can be safely addressed.
We implement periodic boundary conditions and thus can numerically simulate \mbox{Equation (\ref{eqphia})} (with the expressions in Equation (\ref{lapl1})) by using a Fourier approach. This hypothesis is actually mainly justified for what concerns the $w$ direction, which reflects the toroidal symmetry of a tokamak machine. In this sense, we consider the following Fourier expansion: 
\begin{align}
\Phi (u,v,w) = \sum_{n,\ell,m}\varphi_{n,\ell,m}\;e^{i(nw+\ell u+mv)}\,,\label{ccm8}
\end{align}
where the mode numbers $(n,l,m)$ are negative or positive integers, the reality condition of the field $\varphi$ reads $\varphi_{-n,\ell,m} = \varphi_{n,\ell,m}^*$ and $\varphi_{0,-\ell,-m} = \varphi_{0,\ell,m}^*$. Moreover, the mode numbers are  bounded above to satisfy the natural turbulence cut-off related to the ion Larmor radius scale \cite{montani-physicad-2022}, thus resulting in a truncated Fourier expansion. In order to better underline the spectral dynamics of the system without the relevant damping effects, we neglect the ion viscosity by considering $\alpha_2=0$ in the numerical analysis of this section. Equation (\ref{eqphia}) is thus rewritten as
\begin{align}\label{eqpseuspectr}
\partial_\tau\varphi_{n,\ell,m}=\frac{\alpha_1}{\ell^2+m^2}\;\Theta_{n,\ell,m}-n^2\Big(\bar{\alpha}_3+\frac{\bar{\alpha}_4}{\ell^2+m^2}\Big)\varphi_{n,\ell,m}\,,   
\end{align}
with the scaled constants defined as $\bar{\alpha}_{3,4}=\alpha_{3,4} (L_p/L_t)^2$. Here, we denote with $\Theta_{n,\ell,m}$ the 3D convolution (Cauchy product) of the nonlinear term in Equation (\ref{eqphia}). In the numerical simulations, it will be implemented using a pseudo-spectral approach. (The explicit expression can be easily obtained by applying the Fourier expansion.) This consists of resolving the nonlinearity of the system (i.e., the products of evolving variables) directly in the physical space by using inverse fast Fourier transform algorithms. After the product is evaluated in the $(u,v,w)$ space, it is moved back to the spectral space using the direct Fourier transform. This procedure leads to a relevant reduction in the computational time with respect to the direct implementation of the 3D convolution in the $(n,\ell,m)$ space. The anti-aliasing 2/3 technique (zero padding) is also implemented. This technique ensures the validity of the pseudo-spectral approach by removing the issue due to the fact that solving a product in the physical space may induce corresponding sums of modes which fall outside the considered spectral domain. For the time evolution of the components $\varphi_{n,\ell,m}$, a fourth-order Runge--Kutta algorithm was chosen. The time step has been set by ensuring the ``energy'' conservation (of $\mathcal{O}(10^{-8}$)) of the $n=0$ mode.

\begin{figure}[ht!]
\includegraphics[width=5.3cm]{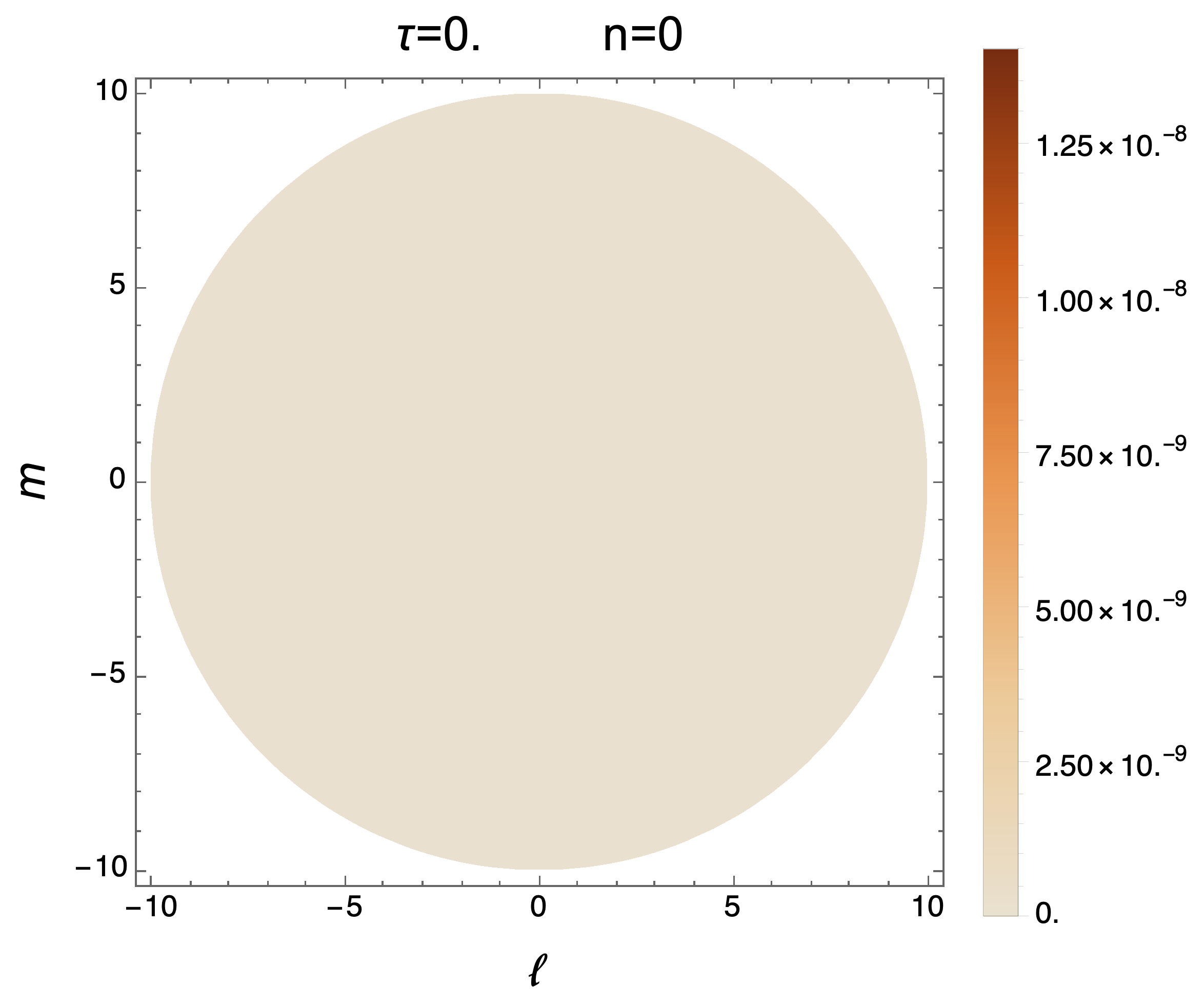}\!\!
\includegraphics[width=5.3cm]{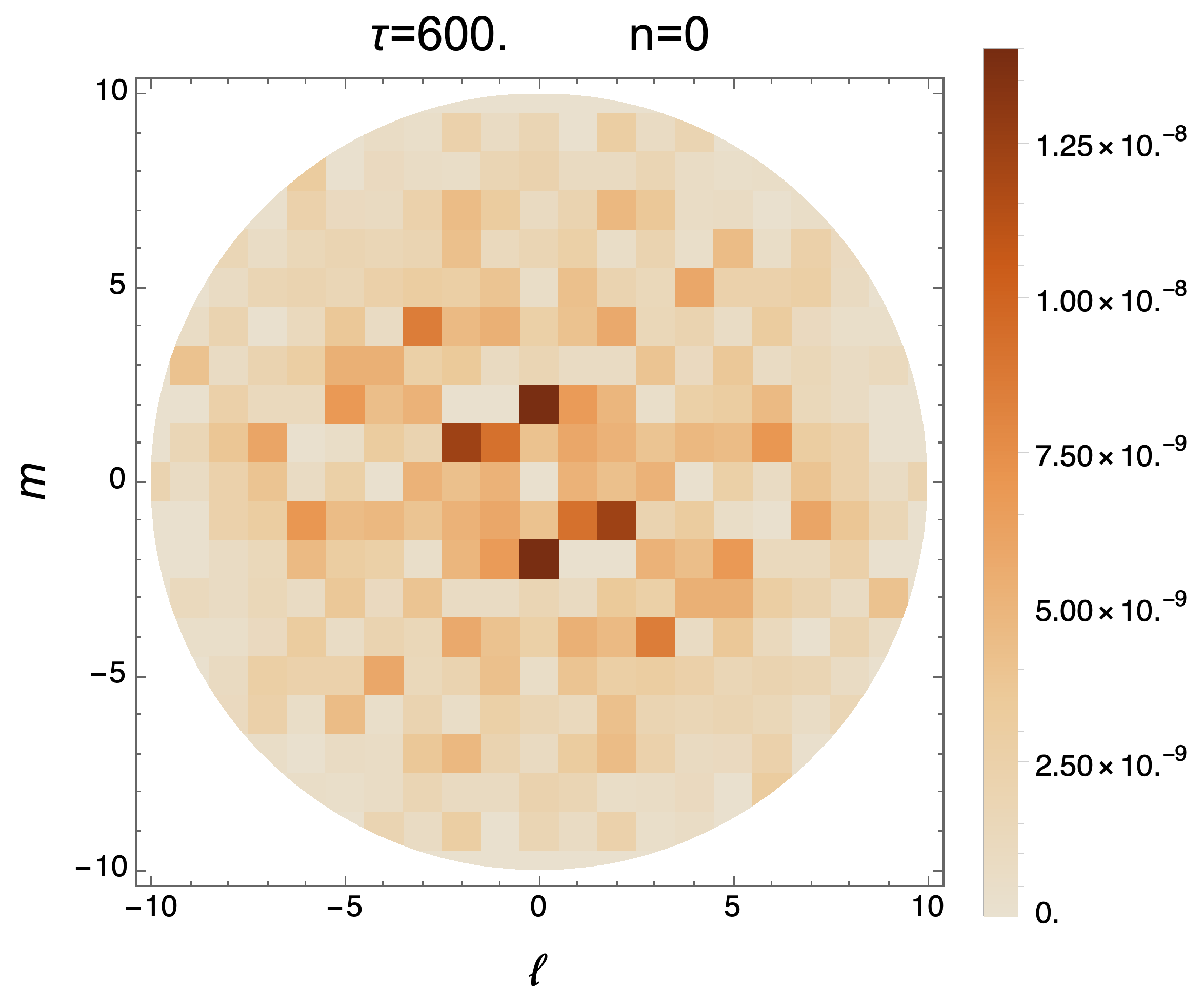}\\
\includegraphics[width=5.3cm]{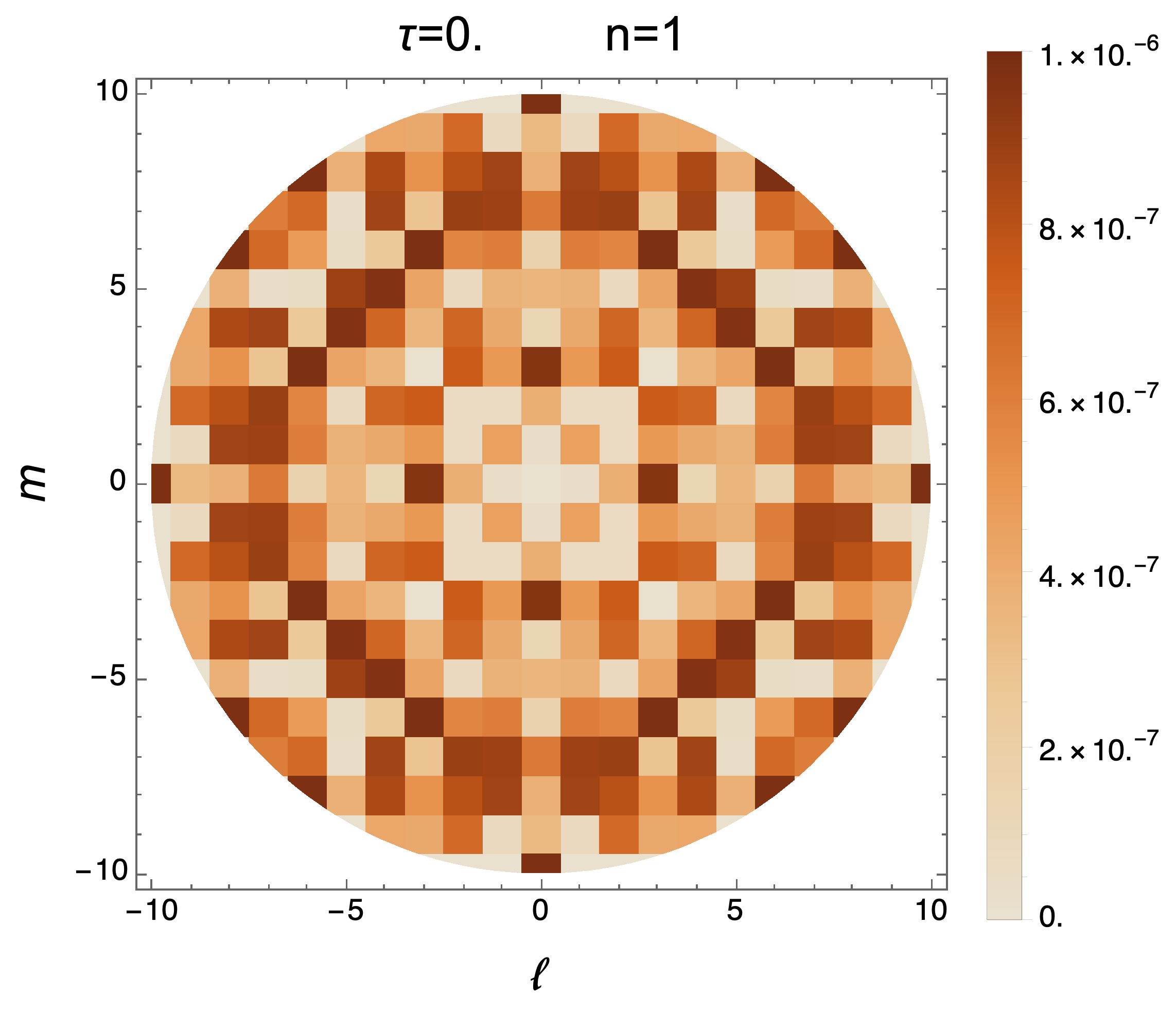}\!\!
\includegraphics[width=5.3cm]{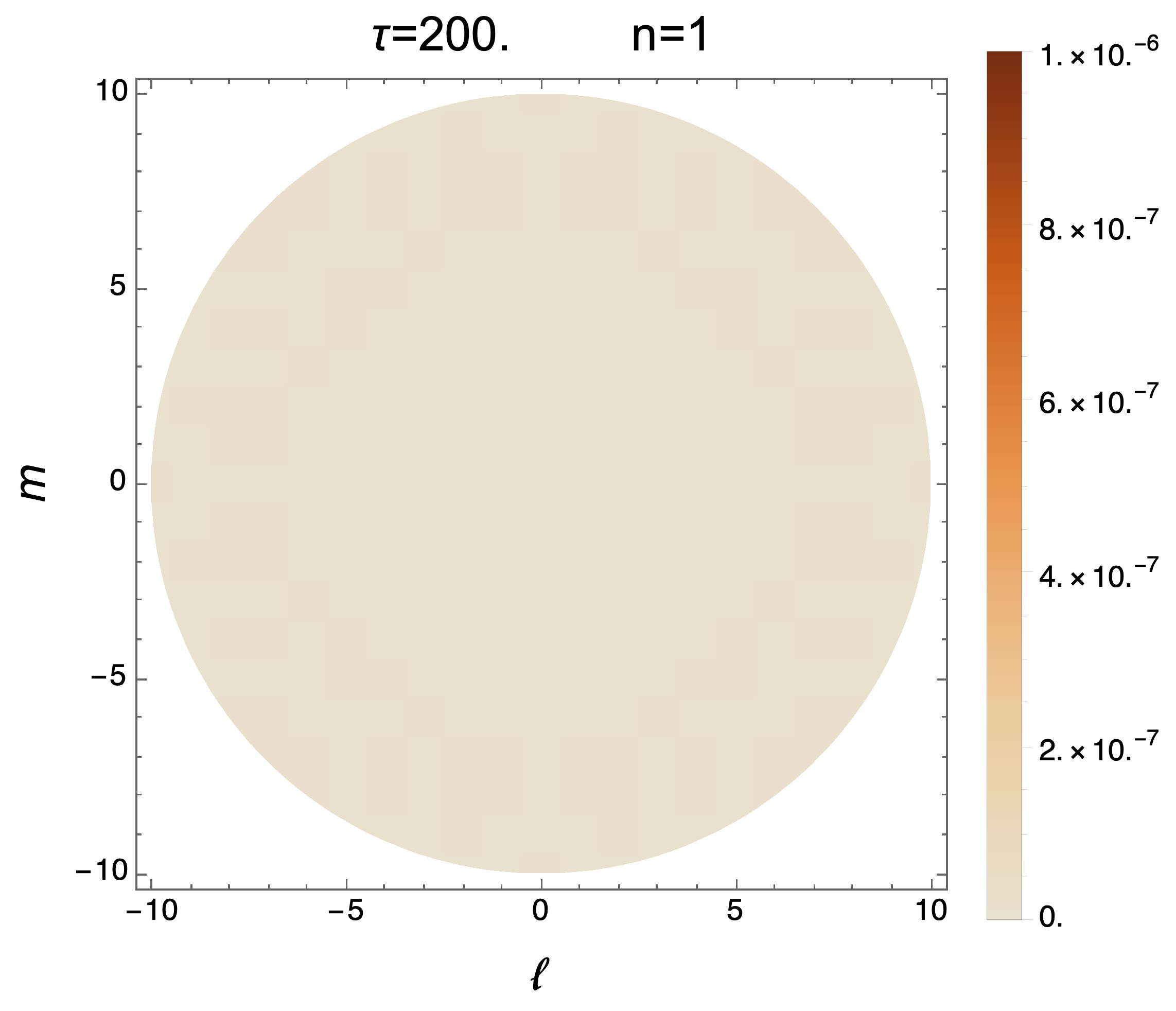}\\
\includegraphics[width=5.3cm]{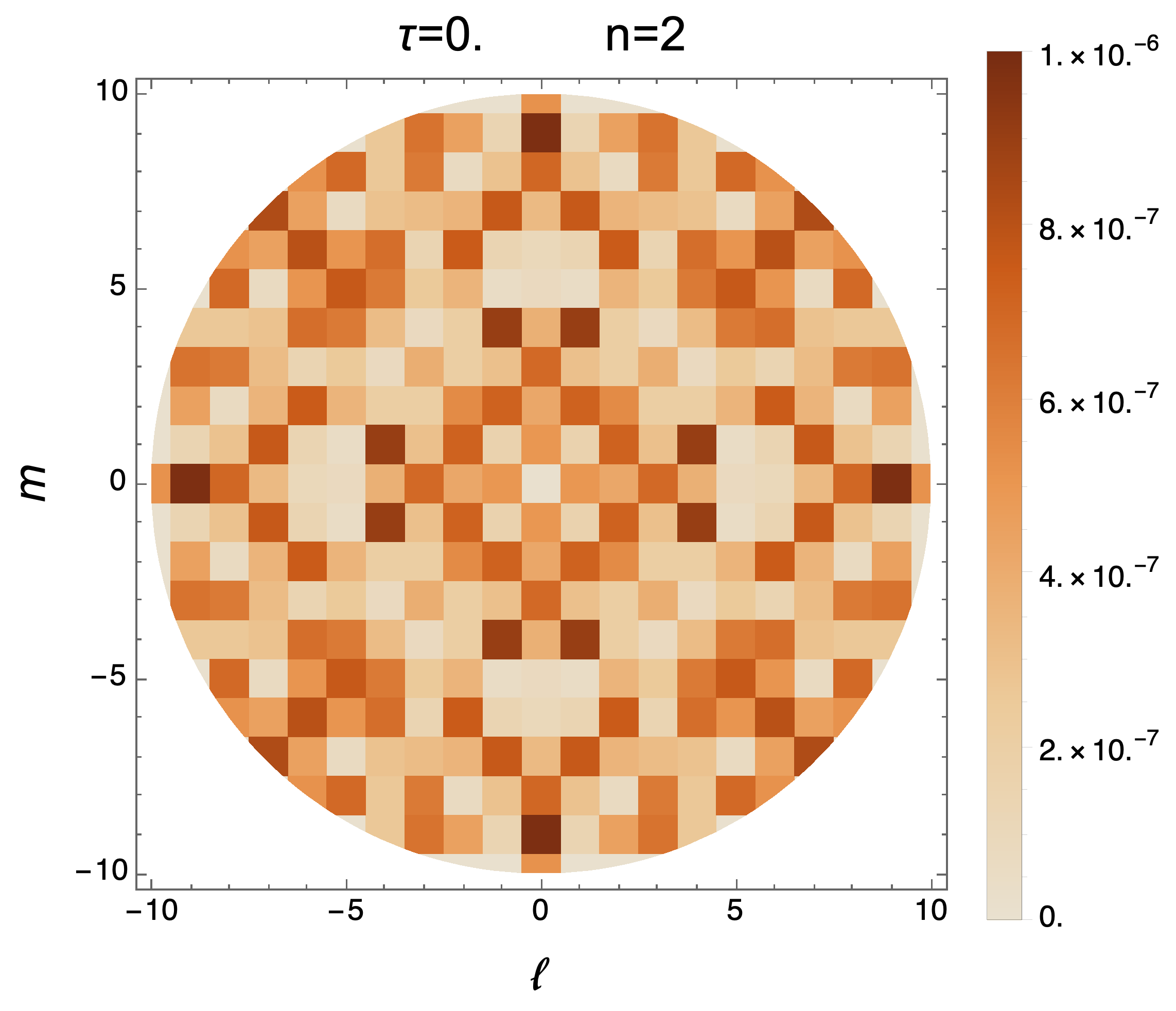}\!\!
\includegraphics[width=5.3cm]{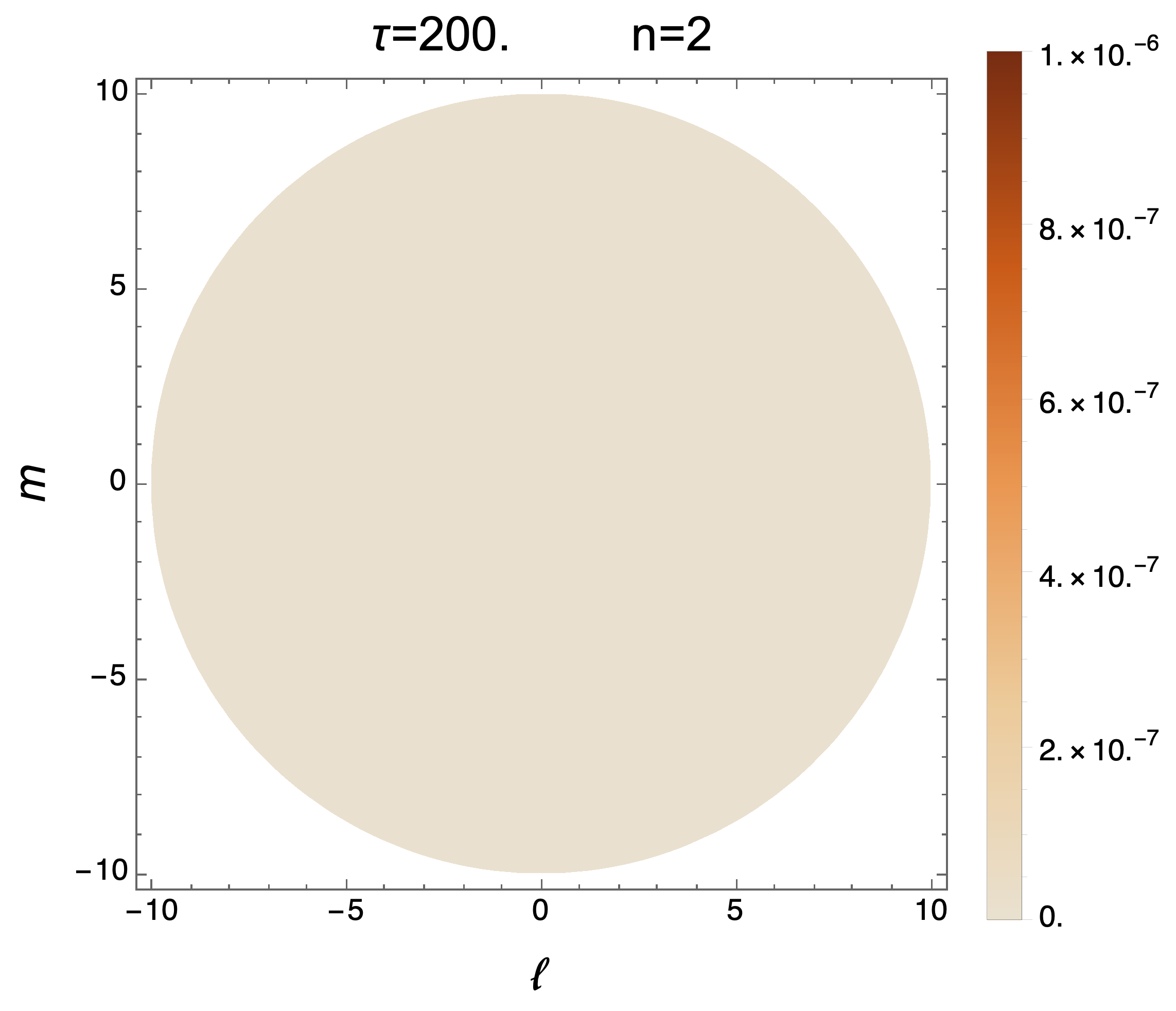}
\caption{Contour plot of $|\varphi_{n,\ell,m}|$ (arbitrary units) for $n=0,1,2$ at different $\tau$, as denoted over the graphs. Data are provided by integrating Equation (\ref{eqpseuspectr}) in the truncated Fourier space with the spatial scale cut off at $2\rho_i$ (color on line).
\label{fig3D}}
\end{figure}

\begin{figure}[ht!]
\includegraphics[width=5.3cm]{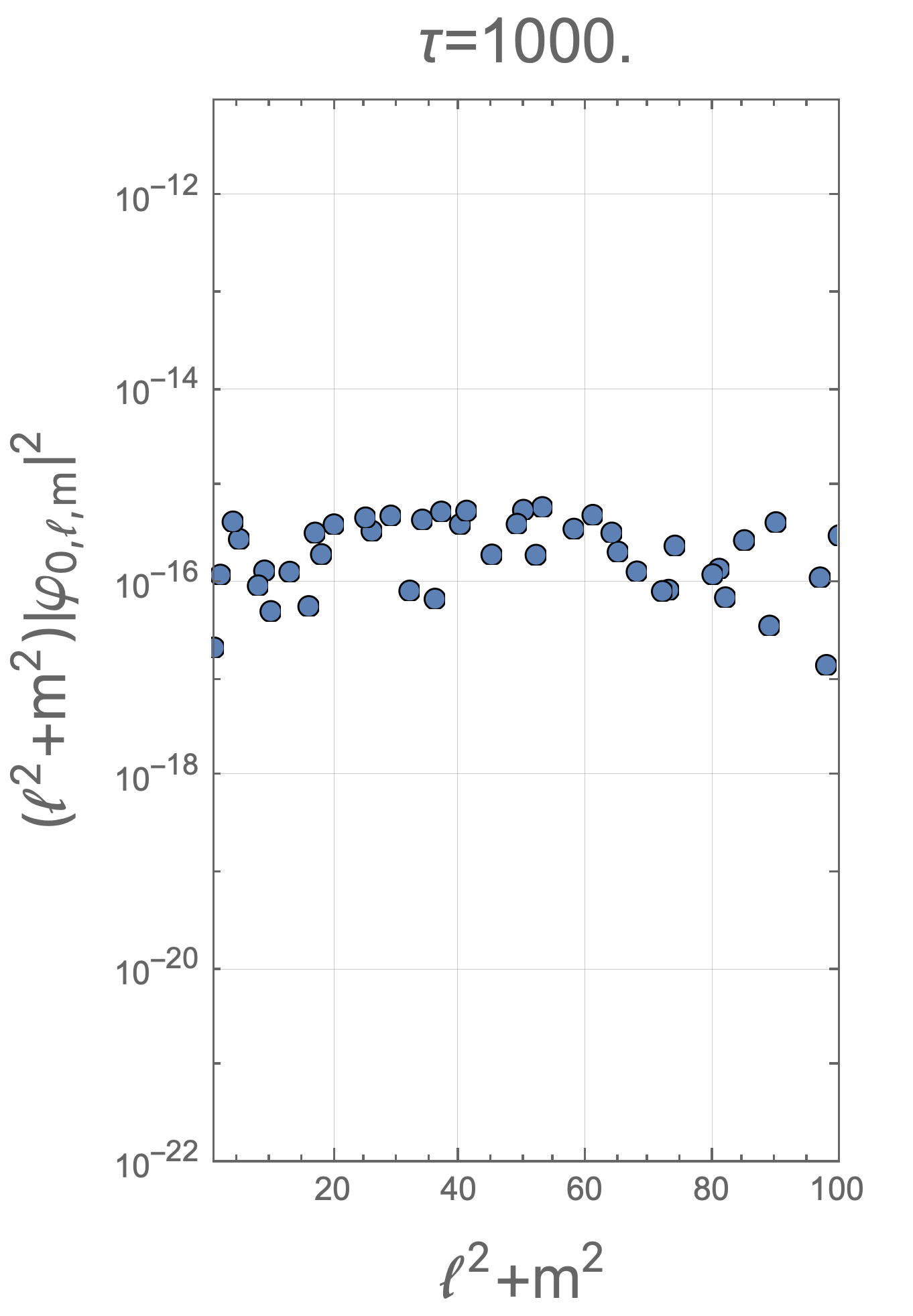}
\caption{Plot (in log scale) of the instantaneous spectrum taken after thermal equilibrium at $\tau=1000$.
\label{figspct}}
\end{figure}

We recall that by ruling out the dependence on $w$ in the whole scheme, the model is actually isomorphic to a 2D incompressible fluid \cite{montani-fluids2022}, where the electric fluctuations correspond to the stream functions \cite{kraichnan80}. The relevant difference between the two schemes relies in the low spatial scale cut-off described above. Moreover, in \cite{montani-physicad-2022}, it was shown theoretically and numerically how the general dynamical system described by \mbox{Equation (\ref{eqpseuspectr})} actually collapses to an axially symmetric structure if the problem is initialized with a dominant contribution from the component $\varphi_{0,\ell,m}$. In fact, the governing equations linearized around $\varphi_{0,\ell,m}\gg\varphi_{n\neq0,\ell,m}$ lead to the constant vorticity solution $\varphi_{0,\ell,m}\propto 1/(\ell^2+m^2)$ for the dominant component and to a time decaying solution for $\varphi_{n\neq0,\ell,m}$. In this section, we point out how this feature is actually a global property of the dynamical system described by Equation (\ref{eqpseuspectr}), since the axial symmetry represents a general attractor of the dynamics even when the $\varphi_{0,\ell,m}$ initial contribution is negligible. 

For this purpose, let us now initialize the simulations of Equation (\ref{eqpseuspectr}) with a random noise $n\neq0$ contribution while the $\varphi_{0,\ell,m}$ components are set zero. Typical tokamak parameters were implemented: $T=100$ eV, $B_t=3$ T and $\mathcal{N}=5\times10^{19}$ m$^{-3}$ \cite{dtt19}. Thus, $\Omega_i\simeq1.4\times 10^8$ s$^{-1}$ and $\rho_i\simeq0.048$ cm. In the poloidal plane, we considered a periodicity box $L_p=1$ cm in length, while $L_t\simeq1300$ cm. The physical cut-off for the small spatial scales was safely set to $2\rho_i$, and for mere numerical reasons, but without losing the generality of the results, we address only $n=-5,\,...\,,5$. We remark that, as discussed above, the choice of a small (compared with typical divertor legs to the order of 30 cm) poloidal box was due to the assumption of treating a pure toroidal magnetic field. This implies a boundary threshold for the small poloidal mode numbers. Actually, due to the relevant cut-off threshold related to the Larmor radius for the large mode numbers, the box size did not quantitatively affect the physical results, since only a few more modes would be taken into account in the dynamics.

In Figure \ref{fig3D}, the contour plot in arbitrary units of $|\varphi_{n,\ell,m}|$ are presented. We considered the initial states of the simulation and subsequent fixed times indicated over the plots (please note the different time scales). As a result, the $n=0$ component is shown to be excited although not initialized, while the $n\neq0$ fluctuations exhibited a decaying behavior (only $n=1,\,2$ are plotted, since the other $n\neq0$ modes behaved accordingly). The mode $n=0$ was the only surviving component of the turbulent dynamics due to an energy transfer from the higher $n$ components. Thus, it can be argued how the axial symmetry scheme emerged as an attractor of the turbulent regime. In this sense, the dynamics was reduced to a pure 2D model in a rather short time compared with the achievement of thermal equilibrium (considered here as a given state for which no deviations of the spectrum emerged, letting the system evolve over time). The ``energy'' spectrum morphology of the $\varphi_{0,\ell,m}$ component is shown in Figure \ref{figspct}. The quantity $(\ell^2+m^2)|\varphi_{0,\ell,m}|^2$ was plotted versus the poloidal wave number (averages over equal  $(\ell^2+m^2)$ values were implemented) at a fixed time taken after thermal equilibrium was reached. The outlined profile can be reasonably considered random but bounded above by a constant value. It is well known from 2D turbulence studies, starting from the pioneering works \cite{seyler75,kraichnan80}, that the equilibrium energy spectral profiles strongly depend on the given initial conditions, and they are reached through energy and enstrophy cascades (the two conserved quantities of the inviscid 2D picture). Specifically, by implementing the canonical ensemble statistics, it is possible to outline a behavior $\propto1/(a_1+a_2(\ell^2+m^2))$ of the 2D energy spectra. (Here, $a_{1,2}$ indicate two constant inverse ``temperatures'' associated with the constant of motion.) We can thus argue how, in the 3D dynamics analyzed in this work, the $n\neq0$ components play the role of a trigger for the axisymmetric instability and then rule out the dynamics due to the decaying behavior. In this sense, the obtained spectrum for the $n=0$ modes corresponds to an initialized 2D profile,  such as $a_1\gg a_2$. A study of the effects of different initial conditions of the $n\neq0$ components on the related spectra is not in the scope of the present paper and should be developed for future works.

\section{Linearized Theory and the Role of X-Point Symmetries}\label{seclin}
In this section, we are interested in characterizing the role of the geometry dependence induced by the poloidal field on the turbulent dynamics in the proximity of the X-point. To this aim, we restrict our analysis to the axisymmetric scheme, which in the previous section was shown to be the only surviving contribution. We can thus focus our analysis on the 2D dynamics only (i.e., on the $(u,v)$ plane representing a poloidal section), hence neglecting the dependence of the dynamical quantities on the toroidal coordinate $w$. By considering a small region of a size $L_p$ around the X-point, we can safely retain $\epsilon\ll1$. In this way, we can treat the contribution given by the poloidal field as a small perturbation. By expanding the parallel gradient $\textbf{D}_\parallel$ and the directional versor $\!\!\hat{\,\,\textbf{b}}$ at lowest order in $\epsilon$, we obtain
\begin{equation}
	\textbf{D}_{\parallel} \to \epsilon 
	\left( v\partial_u + u\partial_v\right)	 \!\!\hat{\,\,\textbf{b}}\,,\qquad
 \!\!\hat{\,\,\textbf{b}}\to \epsilon v \hat{\textbf{e}}_x + \epsilon u\hat{\textbf{e}}_y + \hat{\textbf{e}}_z\,.
	\label{tt3}
\end{equation}
It is now immediate to recast the Laplacian operators in the following form:
\begin{align}
    &D_{\parallel}^2 
	\to \epsilon^2\left( 
	v^2\partial_u^2 + 
	u^2\partial_v^2 
	+ 2u v\partial_u\partial_v + u\partial_u + v\partial_v
	\right)\,,\label{dp1}\\
     &D_{\perp}^2 
	\to \partial_u^2+\partial_v^2-\epsilon^2\left( 
	v^2\partial_u^2 + 
	u^2\partial_v^2 
	+ 2u v\partial_u\partial_v 
	\right)\,,\label{do1}
\end{align} 
 where it can be noticed that the correction due to the poloidal field is $\mathcal{O}\leri{\epsilon^2}$, while the $\mathcal{O}\leri{\epsilon^4}$ terms have been dropped.

In order to gain insight on the problem through analytical techniques, we now investigate the linearized version of the governing equations. As can be noticed from \mbox{Equations (\ref{eq1}) and (\ref{eq2}),} the only nonlinearity of the system comes from the Lagrangian derivative applied to the electric potential fluctuations. Thus, in the following, we approximate
\begin{align}
\frac{d}{d\tau}\simeq \partial_\tau\,.
\end{align}
Moreover, in light of the perturbative approach we are implementing, we assume neglecting the background magnetic field gradients according to the drift ordering scheme (i.e., the fluctuation of the second gradients lead the dynamics). This can easily be seen directly from the first term of  Equation (\ref{lm18}), where it can be noticed how the magnetic field gradients would be multiplied by first-order derivatives in the fluctuating field. These terms are safely considered to be of a higher order with respect to the corresponding second derivatives of the fluctuations. These considerations can also be applied to the second term, provided that the viscous parameter is sufficiently small. Equations (\ref{eq1}) and (\ref{eq2}) are thus simplified as follows:
\begin{align}
\partial_\tau\bar{\mathcal{N}}-\bar{\mathcal{D}} D^2_{\perp} \bar{\mathcal{N}} &= 
\boldsymbol{D}_{\parallel}\cdot\textbf{Y}_{\parallel}\;,\label{eq1b}\\
\alpha_1 \partial_\tau D_\perp^2\Phi
-\alpha_1\alpha_2 D_\perp^4\Phi= \gamma^2 
\boldsymbol{D}_{\parallel}\cdot\textbf{Y}_{\parallel}&\simeq
\boldsymbol{D}_{\parallel}\cdot\textbf{Y}_{\parallel}\;,\label{eq2b}
\end{align}
where, for the approximation in Equation (\ref{eq2b}), we expanded the parameter $\gamma$ for small $\epsilon$ values and ruled out $\mathcal{O}(\epsilon^4)$ by means of Equation (\ref{dp1}). It is now easy to recognize that in this simplified 2D scheme as well, the dynamics can be reduced to a single equation for the electric potential vorticity. In fact, the constitutive relation in Equation (\ref{nphi}) is still valid here (considering $\mathcal{D}=\nu$), and the dynamics of the electric potential is governed by
\begin{equation}
\partial_{\tau}D^2_\perp\Phi
= \alpha_2 D^4_\perp\Phi+\alpha_3 D^2_\parallel D^2_\perp \Phi-\alpha_4 D^2_\parallel\Phi\,.
\label{ittxl}
\end{equation}

This equation corresponds to the 2D linearization at $\mathcal{O}(\epsilon^2)$ of the dynamical system in Equations (\ref{eq1}--\ref{lm19}) (we recall that $\alpha_4=\alpha_3/\alpha_1$). Its form is clearly equal to that of Equation (\ref{eqphia}), apart from the nonlinear term, but the Laplacian operators are now expressed in Equations (\ref{dp1}) and (\ref{do1}). It can be stressed how the X-point geometry (particularly the presence of a small but nonzero poloidal field) introduces a parallel gradient, taken here at $\mathcal{O}(\epsilon^2)$, which naturally emerges in the 3D modeling. This approximation leads to a good description of the dynamics as long as the quadratic terms in the vorticity are much smaller than the linear terms retained here.

In order to calculate the regular solutions of the dynamics, we followed a perturbative approach. Specifically, we are interested in characterizing the role of the X-point geometry, which is treated as a small contribution able to slightly modify the profile of the background solution obtained by setting the small parameter $\epsilon=0$. To this end, we split the electric potential into two contributions:
\begin{equation}
	\Phi = \Phi^{(0)}(\tau,u,v) + 
	\epsilon^2 \Phi^{(1)}(\tau,u,v)\, , 
	\label{sx1}
\end{equation}
and we additionally require that  $\epsilon^2 |\Phi^{(1)}|\ll |\Phi^{(0)}|$ for all times. By substituting and separating the zero and first order in $\epsilon^2$, we obtain the following two equations,
where the first one determines the $\Phi^{(0)}$ term (i.e., the background solution corresponding to the absence of any contribution from the X-point geometry):
\begin{equation}
	\partial_{\tau}(\partial_u^2+\partial_v^2)\Phi^{(0)} -\alpha_2 (\partial_u^2+\partial_v^2)^2\Phi^{(0)}=0.
	\, 
	\label{sx2}
\end{equation}
Meanwhile, the second one
\begin{multline}
	\partial_{\tau}(\partial_u^2+\partial_v^2)\Phi^{(1)} -\alpha_2 (\partial_u^2+\partial_v^2)^2\Phi^{(1)} =
	 \partial_{\tau}\left( v^2\partial_u^2 + 
	u^2\partial_v^2 
	+ 2u v\partial_u\partial_v \right) \Phi^{(0)} -\\
 -2\alpha_2 \Big( v^2\partial_u^4 + 
u^2\partial_v^4+\leri{u^2+v^2} \partial_u^2\partial_v^2
	+ 2u v\partial_u\partial_v \leri{\partial_u^2+\partial_v^2}+\\
\qquad + 4u\partial_u\partial^2_v+4v\partial_v\partial^2_u+\partial_u^2+\partial_v^2 \Big)\Phi^{(0)}- \\
 -\alpha_4 \leri{v^2\partial_u^2 + 
	u^2\partial_v^2 
	+ 2u v\partial_u\partial_v + u\partial_u + v\partial_v}\Phi^{(0)}+\\
 +\alpha_3 \leri{v^2\partial_u^2 + 
	u^2\partial_v^2 
	+ 2u v\partial_u\partial_v + u\partial_u + v\partial_v}\leri{\partial_u^2+\partial_v^2}\Phi^{(0)}
	\, ,
	\label{sx3}
\end{multline}
provides the profile of the perturbation $\Phi^{(1)}$, which is the first-order contribution given by the adopted magnetic configuration displayed in Equation (\ref{lm}).
We selected the solution of the zero-order problem in Equation \eqref{sx2}, which is nothing more than a heat equation, satisfying the Dirichlet boundary conditions. Without loss of generality, we set such a solution to be 
\begin{equation}
    \Phi^{(0)}(\tau,u,v)=e^{-\alpha_2(\ell_0^2+m_0^2)\tau} \sin(\ell_0 u) \sin(m_0 v)
\end{equation}
with $(\ell_0,m_0)$ as two positive integers. By inserting this explicit form into Equation \eqref{sx3}, we obtain for the perturbation $\Phi^{(1)}$ 
\begin{equation}\label{eqphi1}
    \partial_{\tau}(\partial_u^2+\partial_v^2)\Phi^{(1)} -\alpha_2 (\partial_u^2+\partial_v^2)^2\Phi^{(1)}= e^{-\alpha_2(\ell_0^2+m_0^2)\tau} \, F(u,v) ,
\end{equation}
where we have introduced 
\begin{multline}
F(u,v) \equiv 2\ell_0 m_0 uv\leri{\leri{\alpha_2-\alpha_3}\leri{\ell_0^2+m_0^2}-\alpha_4}\cos(\ell_0 u) \cos(m_0 v)+\\
+\ell_0 u \leri{8\alpha_2 m_0^2-\alpha_4-\leri{\ell_0^2+m_0^2}\alpha_3}\cos(\ell_0 u)\sin(m_0v)+\qquad\qquad\qquad\qquad\\
+m_0v \leri{8\alpha_2 \ell_0^2-\alpha_4-\leri{\ell_0^2+m_0^2}\alpha_3}\sin(\ell_0 u)\cos(m_0v)+\qquad\qquad\qquad\qquad\\
+\bigg(\leri{\leri{\alpha_3-\alpha_2}\leri{\ell_0^2+m_0^2}+\alpha_4}\leri{m_0^2u^2+\ell_0^2v^2}+2\alpha_2\leri{\ell_0^2+m_0^2}\bigg)\sin(\ell_0u)\sin(m_0v).
\end{multline}

It can be noticed that even when starting from a background solution $\Phi^{(0)}$ showing the property of being null on the square boundaries, we obtained a source term $F(u,v)$ for which
\begin{align}
&F(0,v)=0 \qquad F(2\pi,v) \neq 0\\
&F(u,0)=0 \qquad F(u,2\pi) \neq 0.
\end{align}

We solved Equation \eqref{eqphi1} by expanding the spatial part of $\Phi^{(1)}$ in the Fourier series and performing a Laplace transform on the time coordinate $\tau$. For what concerns the treatment of the spatial part, we will apply the 2D version of the Fourier expansion displayed in Equation \eqref{ccm8}. For completeness, we report the formula through which we calculated the mode amplitudes in the Fourier space for a generic real-valued function of the spatial coordinates, namely
\begin{equation}\label{amplitudes}
f_{\ell,m}=\dfrac{1}{4\pi^2}\int_0^{2\pi}  du \int_0^{2\pi}   dv \, e^{-i\leri{\ell u+mv}}f(u,v),
\end{equation}
where $(\ell,m)$ are indices running on the entire set of integer numbers. The Laplace transform of a real-valued function is defined as
\begin{equation}
g_s=\int_0^\infty d\tau \, e^{-s\tau} g(\tau)\,,\qquad
g(\tau)=\dfrac{1}{2\pi i} \int_P ds \, e^{s \tau} g_s\,,
\end{equation}
with $s\in \mathbb{C}$ and $P$ being a path in the complex $s$ plane such that any $s\in P$ results in $\Re (s)>\Re(s_k)$, where $s_k$ represents the poles of the Laplace transform $g_s$. We proceed by applying the aforementioned transform on both sides of Equation \eqref{eqphi1} so that a solution can be calculated through simple algebraic manipulations, resulting in
\begin{equation}\label{phi1spq}
\Phi^{(1)}_{s,\ell,m}=\dfrac{\Phi^{(1)}_{\ell,m}(0)}{s+\alpha_2 \leri{\ell^2+m^2}}-\dfrac{F_{\ell,m}}{\leri{\ell^2+m^2}\leri{s+\alpha_2 \leri{\ell^2+m^2}}\leri{s+\alpha_2 \leri{\ell_0^2+m_0^2}}},
\end{equation}
where $\Phi^{(1)}_{\ell,m}(0)$ and $F_{\ell,m}$ are the Fourier amplitudes of the initial datum $\Phi^{(1)}(0,u,v)$ and the external source $F(u,v)$, respectively. (The explicit forms of $F_{\ell,m}$ can be found in \mbox{Appendix A.)}  Hence, the perturbation $\Phi^{(1)}$ in the coordinate space reads as follows:
\begin{equation}
    \Phi^{(1)}(\tau,u,v)=\dfrac{1}{2\pi i} \int_P ds \, e^{s \tau}\sum_{\ell,m}  e^{i\leri{\ell u+mv}}  \Phi^{(1)}_{s,\ell,m}.
\end{equation}

Now, a simple consideration holds: When evaluating the inverse Laplace transform, we noticed that the term containing the initial datum $\Phi^{(1)}_{\ell,m}(0)$ was characterized by a single pole located at $s=-\alpha_2 \leri{\ell^2+m^2}$, whereas the term related to the external source $F_{\ell,m}$ featured a supplemental pole in $s=-\alpha_2 \leri{\ell_0^2+m_0^2}$, as can easily be inferred from the inspection of Equation \eqref{phi1spq}. In both cases, we dealt with poles located on the negative real semi-axis, and the path of integration $P$ could be chosen to be coincident with the imaginary axis (i.e. $s=i \omega$, with $\omega$ being a real variable ranging across all real numbers):
\begin{align}
\Phi^{(1)}(\tau,u,v)=\dfrac{1}{2\pi}\sum_{\ell,m} e^{i\leri{\ell u+mv}} 
     \int_{\mathbb{R}}
     d\omega \, \dfrac{e^{i \omega \tau}}{i\leri{\omega-i\alpha_2 \leri{\ell^2+m^2}}}\lerisq{ \Phi^{(1)}_{\ell,m}(0)+\dfrac{iF_{\ell,m}}{\leri{\ell^2+m^2}\leri{\omega-i\alpha_2 \leri{\ell_0^2+m_0^2}}}}.
\end{align}

\begin{figure}[ht!]
\includegraphics[width=6cm]{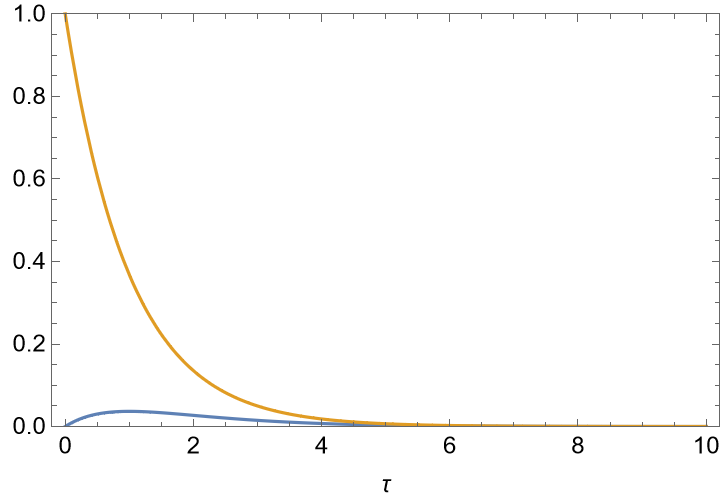}\;
\includegraphics[width=6cm]{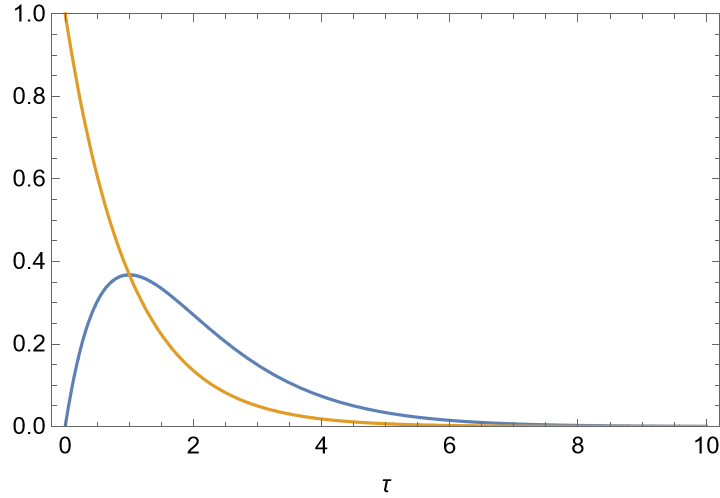}
\caption{Time evolution of $\Phi^{(0)}$ (orange) and $\epsilon^2 \Phi^{(1)}$ (blue) in arbitrary units. The parameter $\tau_{EXP}$ is set to $1$ in both cases, whereas $\tau_{RES}$ is set to $10$ and $1$ in the left and right panels, respectively (color online).}
\label{figres}
\end{figure}

Let us now focus on the second term inside the square brackets. When calculating the integral through the application of the residue theorem, it must be taken into account that the two poles characterizing this contribution are distinguished by $\ell^2+m^2\neq \ell_0^2+m_0^2$, whereas a double pole appears when $\ell^2+m^2=\ell_0^2+m_0^2$ holds. To be more specific, in the first case, when we calculated the perturbation $\Phi^{(1)}$ in the coordinate space through an inverse transform, we dealt with an integral of the form 
\begin{equation}
    \mathcal{I}_1=\int_{\mathbb{R}}
     d\omega \, \dfrac{e^{i  \omega \tau}}{\omega-i \omega_0} \,,
\end{equation}
with $\omega_0>0$ being a positive real constant and $t$ being a variable ranging across the positive real semi-axis. The residue theorem allows an immediate evaluation of this quantity, resulting in $\mathcal{I}_1= 2\pi i \, e^{-\omega_0 \tau}$. On the other hand, when the values of the Fourier momenta $\ell$ and $m$ are such that $\ell^2+m^2=\ell_0^2+m_0^2$, we have to treat an integral characterized by a double pole, namely
\begin{equation}
\mathcal{I}_2=\int_{\mathbb{R}}
     d\omega \, \dfrac{e^{i  \omega \tau}}{\leri{\omega-i \omega_0}^2}\,,
\end{equation}
in which case the application of the residue theorem returns $\mathcal{I}_2=- 2\pi \tau \, e^{-\omega_0 \tau}$. Therefore, the temporal dependence of all the terms for which $\ell^2+m^2\neq \ell_0^2+m_0^2$ will be $e^{-\alpha_2 k^2 \tau}$, where we denote with $k^2$ any possible combination of momenta. On the contrary, for all the terms satisfying $\ell^2+m^2= \ell_0^2+m_0^2$ and giving rise to the presence of a double pole, we calculated a temporal dependence of the form $\tau e^{-\alpha_2\leri{\ell_0^2+m_0^2}\tau}$. The arising of these latter terms is due to a resonance phenomenon between the perturbation and the background solution. Indeed, it is well known that the introduction in a differential equation of an external force having a frequency close or coincident with one of the natural frequencies of the associated homogeneous equation causes the presence of terms increasing the overall amplitude on specific time scales. In our case, the secular growth of the resonant terms took place on a time scale $\tau_{RES}=1/\epsilon^2$, whereas the exponential term $e^{-\alpha_2 \leri{\ell_0^2+m_0^2} \tau}$ became dominant for times of the order $\tau_{EXP}=1/(\alpha_2 \leri{\ell_0^2+m_0^2})$. Therefore, it is straightforward to recognize that for $\tau_{RES}\gg \tau_{EXP}$, we obtained a regular perturbative expansion satisfying $\epsilon^2| \Phi^{(1)}|\ll| \Phi^{(0)}|$ for all times. Conversely, if $\tau_{RES}\lesssim \tau_{EXP}$, then we noticed the appearance of a region of times in which the ansatz underlying the expansion in Equation \eqref{sx1} was violated. The two different behaviors of the solutions just described are shown in  Figure \ref{figres}. Then, we claim that for $\tau_{RES}\lesssim \tau_{EXP}$, Equation \eqref{ittxl} must be investigated without resorting to a perturbative approach. 
Let us now be more specific and apply these findings to a concrete and realistic scenario. Using the typical tokamak parameters introduced in the previous section, for the viscosity constant, we obtained $\alpha_2\simeq6.84\times10^{-6}$ from Equation (\ref{visc}). At the same time, the relevant ratio of the poloidal to toroidal magnetic field can be deduced from the linear scaling in a region close to the X-point. For instance, by considering a typical equilibrium configuration (in the presence of a bottom single null), we found the ratio $\simeq10^{-2.1}$ in a region of $\sim$3 cm around the X-point \cite{DTT_2021}, and thus we can assume $\epsilon=4.21\times10^{-4}$ in the poloidal box of a length $L_p=1$ cm we are considering.
The condition for which the perturbative expansion results are valid (i.e. $\tau_{RES}\gg \tau_{EXP}$) can be recast as $\ell_0^2+m_0^2 \gg \epsilon^2/\alpha_2=2.6\times 10^{-2}$, and it is satisfied by any $(\ell_0,m_0)$ in the set of positive integers. However, the ratio $\epsilon^2/\alpha_2$ is extremely sensitive to the particular choice of the poloidal box size. Indeed, the dependence on this parameter is $\epsilon^2/\alpha_2 \propto L_p^4$, and it is sufficient to slightly enlarge the considered region around the X-point to exclude certain values for $(\ell_0,m_0)$. For instance, by taking $L_p=5$ cm, we obtained the constraint $\ell_0^2+m_0^2 \gg 16$, which is clearly not satisfied by $(\ell_0,m_0)=(1,1),(2,1),(1,2)\dots (4,2), (2,4)$. For these low-wavenumber modes, the perturbative expansion introduced in Equation \eqref{sx1} is not feasible, and one can gain insight on the behavior of such solutions only through a direct inspection of Equation \eqref{ittxl}.

\section{Concluding Remarks}

We analyzed a reduced model for the turbulence 
in a tokamak plasma edge near the X-point, based on a drift fluid 
approach to the ion and electron dynamics. 
More specifically, we considered a Hasegawa--Wakatani 
formulation, and then, by neglecting the background density gradient and suitably modeling the particle diffusion, we arrived to writing down a single equation for the electric 
potential field for two specific cases. We deepened the two different but complementary 
aspects of this reduced formulation (i.e., on the
one hand, the relaxation process of the 3D turbulence on the 2D interchange-like nonlinear dynamics and, 
on the other hand, the presence of an 
X-point geometry in the linearized dynamics).

The numerical analysis of the 3D nonlinear drift turbulence clarified that even if we initialized only 
the non-axisymmetric modes, after a given time, 
the only surviving configuration was the 
$n=0$ interchange-like fluctuation, which stresses how 
such 2D physics can constitute a basic ingredient 
for determining the transport properties in the edge plasma turbulence.

The study of an axisymmetric linearized system in the presence 
of the X-point geometry was developed via 
a perturbative approach with the small parameter 
$\epsilon$ and by neglecting, according to the 
drift ordering scenario, the background magnetic 
field gradients. We demonstrated that, being close enough to 
the null configuration, the system remained stable, 
as it would be when the magnetic field was along 
the toroidal direction only, but in the outer regions, 
the perturbation scheme failed due to a secular growth of the 
contribution induced by the poloidal field. 
Thus, the presence of a small poloidal magnetic component 
significantly influenced the global system dynamics, 
preventing consideration of the constant homogeneous axial
field as a dominant component sufficiently far from the 
X-point. 
This result suggests that the presence of 
a null configuration can deeply alter the 
turbulence properties with respect to the 
simplified standard picture, emerging when 
a constant magnetic field lying in the $z$ direction is present. In other words, we could expect that 
some specific features of the so-called 
nonlinear drift response can be significantly affected by the peculiar character of the X-point configuration. 

To summarize, the present study individualizes 
the 2D turbulence in the presence of an X-point 
geometry as the basic ingredient to be accounted for when constructing a reduced model able to capture the 
most important aspects of the outward turbulent 
transport in the tokamak edge. 

\vspace{1cm}
\textbf{Acknowledgement} We would like to thank Giulio Rubino for his valuable suggestions about real parameters for incoming okamak devices.

\section{Appendix A}
In this appendix, we report the explicit form of the amplitudes $F_{\ell,m}$ obtained from the application of Equation \eqref{amplitudes} for any possible value of the Fourier momenta $(\ell,m)$:
\begin{equation}\label{diversi}
    F_{\ell,m}=\dfrac{2 \ell_0 m_0\ell m \leri{\alpha_3-\leri{\ell_0^2+m_0^2}\leri{\alpha_2-\alpha_4}}}{\leri{\ell_0^2-\ell^2}\leri{m_0^2-m^2}}\,, \qquad \ell \neq \pm \ell_0 , m \neq \pm m_0
\end{equation}
\begin{multline}
F_{\ell_0,m}=\dfrac{ m_0 m\leri{4\ell_0^4\leri{\alpha_2-\alpha_4}-\ell_0^2\leri{4\alpha_3+m_0^2\leri{3\alpha_2+4\alpha_4}-7\alpha_2 m^2}+\alpha_2m_0^2\leri{m_0^2-m^2}}}{2\leri{m_0^2-m^2}^2}+\\
 +i\leri{\dfrac{\pi \ell_0 m_0\leri{\ell_0+m}\leri{\alpha_3+\leri{\alpha_4-\alpha_2}\leri{\ell_0^2+m_0^2}}}{m_0^2-m^2}}\,,\qquad m\neq \pm m_0
\end{multline}
\begin{multline}
F_{-\ell_0,m}=-\dfrac{ m_0 m\leri{4\ell_0^4\leri{\alpha_2-\alpha_4}-\ell_0^2\leri{4\alpha_3+m_0^2\leri{3\alpha_2+4\alpha_4}-7\alpha_2m^2}+\alpha_2m_0^2\leri{m_0^2-m^2}}}{2\leri{m_0^2-m^2}^2}+\\
+i\leri{\dfrac{\pi \ell_0 m_0\leri{m-\ell_0}\leri{\alpha_3+\leri{\alpha_4-\alpha_2}\leri{\ell_0^2+m_0^2}}}{m_0^2-m^2}}\,,\qquad m\neq \pm m_0
\end{multline}
\begin{multline}
F_{\ell,m_0}=\dfrac{ \ell_0 \ell\leri{4m_0^4\leri{\alpha_2-\alpha_4}-m_0^2\leri{4\alpha_3+\ell_0^2\leri{3\alpha_2+4\alpha_4}-7\alpha_2\ell^2}+\alpha_2\ell_0^2\leri{\ell_0^2-\ell^2}}}{2\leri{\ell_0^2-\ell^2}^2}+\\
 +i\leri{\dfrac{\pi \ell_0 m_0\leri{m_0+\ell}\leri{\alpha_3+\leri{\alpha_4-\alpha_2}\leri{\ell_0^2+m_0^2}}}{\ell_0^2-\ell^2}}\,,\qquad \ell\neq \pm \ell_0
\end{multline}
\begin{multline}
F_{\ell,-m_0}=-\dfrac{ \ell_0 \ell\leri{4m_0^4\leri{\alpha_2-\alpha_4}-m_0^2\leri{4\alpha_3+\ell_0^2\leri{3\alpha_2+4\alpha_4}-7\alpha_2\ell^2}+\alpha_2\ell_0^2\leri{\ell_0^2-\ell^2}}}{2\leri{\ell_0^2-\ell^2}^2}+\\
+i\leri{\dfrac{\pi \ell_0 m_0\leri{\ell-m_0}\leri{\alpha_3+\leri{\alpha_4-\alpha_2}\leri{\ell_0^2+m_0^2}}}{\ell_0^2-\ell^2}}\,,\qquad \ell\neq \pm \ell_0
\end{multline}
\begin{multline}
F_{\ell_0,m_0}=-\dfrac{\pi^2}{6}\leri{2\ell_0^2+3\ell_0m_0+2m_0^2}\leri{\alpha_3+\leri{\alpha_4-\alpha_2}\leri{\ell_0^2+m_0^2}}+\\
 +\dfrac{1}{8\ell_0^2m_0^2}\leri{2\alpha_2 \ell_0^2m_0^2\leri{\ell_0^2+m_0^2}+\alpha_3\leri{\ell_0^4-\ell_0^2m_0^2+m_0^4}+\leri{\alpha_4-\alpha_2}\leri{\ell_0^6+m_0^6}}+\\
 +\dfrac{i\pi}{4\ell_0 m_0}\leri{\leri{\alpha_2\ell_0 m_0+\alpha_3}\leri{\ell_0^3+m_0^3}+\leri{\alpha_4-8\alpha_2}\ell_0^2m_0^2\leri{\ell_0+m_0}+\leri{\alpha_4-\alpha_2}\leri{\ell_0^5+m_0^5}}\,,
 \label{uguali}
\end{multline}
\begin{multline}\label{uguali2}
F_{-\ell_0,m_0}=\dfrac{\pi^2}{6}\leri{2\ell_0^2-3\ell_0 m_0+2m_0^2}\leri{\alpha_3+\leri{\alpha_4-\alpha_2}\leri{\ell_0^2+m_0^2}}+\\
+\dfrac{1}{8\ell_0^2m_0^2}\leri{2\alpha_2 \ell_0^2m_0^2\leri{\ell_0^2+m_0^2}+\alpha_3\leri{\ell_0^4-\ell_0^2m_0^2+m_0^4}+\leri{\alpha_4-\alpha_2}\leri{\ell_0^6+m_0^6}}+\\
+\dfrac{i\pi}{4\ell_0m_0}\leri{\leri{\alpha_2\ell_0 m_0-\alpha_3}\leri{\ell_0^3-m_0^3}+\leri{\alpha_4-8\alpha_2}\ell_0^2m_0^2\leri{m_0-\ell_0}+\leri{\alpha_4-\alpha_2}\leri{m_0^5-\ell_0^5}}\,.
\end{multline}
The remaining amplitudes can be calculated using the reality condition $ F_{\ell,m}^*=F_{-\ell,-m}$.

\newpage


\end{document}